\documentclass{llncs}
\usepackage{graphicx}
\usepackage{latexsym}
\usepackage{tikz}

\usepackage{url}
\usetikzlibrary{trees}
\usepackage{myalgorithm}
\usepackage{mycode}

\def\ILEXISTE.{\ensuremath{\exists}}
\def\QUELQUESOIT.{\ensuremath{\forall}}
\def\ET.{\ensuremath{\wedge}}
\def\OU.{\ensuremath{\vee}}
\def\TOP.{\ensuremath{\top}}
\def\BOTTOM.{\ensuremath{\bot}}
\def\QUANTIFICATEUR.{\ensuremath{q}}
\def\BINDER.{\ensuremath{Q}}

\newcommand{\qcsp}[5]{\ensuremath{( #1, #2, #3, #4, #5 )}}

\newcommand{\arb}[2]{\ensuremath{\mathcal{T}_{#1}(#2)}}
\newcommand{\Intu}[1]{\ensuremath{I(#1)}}
\newcommand{\Intb}[2]{\ensuremath{I^{#1}(#2)}}

\def\QUANT.{\mbox{quant}}
\newcommand{\quant}[1]{\ensuremath{\QUANT.(#1)}}
\def\RANK.{\mbox{rank}}
\newcommand{\rank}[1]{\ensuremath{\RANK.(#1)}}
\def\COMPDOM.{\ensuremath{\overline{(.)}}}

\def\RCQCSP.{\ensuremath{rec\_comp}}
\newcommand{\rcqcsp}[2]{\ensuremath{\RCQCSP.(#1,#2)}}
\def\IQCSP.{\ensuremath{it\_comp}}

\def\COMBILEXISTE.{\ensuremath{\oplus_{\exists}}}
\newcommand{\combilexiste}[3]{\ensuremath{\COMBILEXISTE.(#1,#2,#3)}}
\def\COMBQUELQUESOIT.{\ensuremath{\oplus_{\forall}}}
\newcommand{\combquelquesoit}[3]{\ensuremath{\COMBQUELQUESOIT.(#1,#2,#3)}}
\def\COMBINE.{\ensuremath{\oplus}}
\newcommand{\combine}[2]{\ensuremath{\COMBINE.(#1,#2)}}

\def\SETVARIABLES.{\mbox{\bf V}}
\def\SETDOMAINES.{\mbox{\bf D}}
\def\SETCONSTRAINTS.{\mbox{\bf C}}

\def\SETCONTRAINTES.{\ensuremath{\mathcal{C}}}

\newcommand{\bl}[2]{\ensuremath{\langle #1 \;|\; #2 \rangle}}

\def\FONC.{\ensuremath{\mapsto}}
\newcommand{\fonc}[2]{\ensuremath{(#1 \FONC. #2)}}

\newcommand{\substxparysq}[2]{\ensuremath{[#1 \leftarrow #2]}}

\def\BLTOP.{\ensuremath{bl\_top}}
\def\BLBOTTOM.{\ensuremath{bl\_bottom}}
\def\HEAD.{\ensuremath{head}}
\def\TAIL.{\ensuremath{tail}}
\def\VIDE.{\ensuremath{empty}}
\newcommand{\head}[1]{\ensuremath{\HEAD.(#1)}}
\newcommand{\tail}[1]{\ensuremath{\TAIL.(#1)}}
\newcommand{\vide}[1]{\ensuremath{\VIDE.(#1)}}
\def\success.{\ensuremath{succes}}
\def\failure.{\ensuremath{failure}}

\begin{document}

\title{Compilation for QCSP}

\author{Igor St\'{e}phan\institute{LERIA, University of Angers, France\\ email: igor.stephan@info.univ-angers.fr} }

\maketitle
\bibliographystyle{plain}

\begin{abstract}
We propose in this article a framework for compilation of quantified constraint satisfaction problems (QCSP).
We establish the semantics of this formalism by an interpretation to a QCSP.
We specify an algorithm to compile a QCSP embedded into a search algorithm and based on the inductive semantics of QCSP.
We introduce an optimality property and demonstrate the optimality of the interpretation of the compiled QCSP.

\end{abstract}

\section{Introduction}
A constraint satisfaction problem (CSP) requires a value, selected from a given finite domain, to be assigned to each
variable in the problem, so that all constraints relating the variables are satisfied~\cite{Mackworth_AI_77,Brailsford_Potts_Smith_EJOR_99}.
A quantified constraint satisfaction problem (QCSP)~\cite{Chen_PhD_04,Bordeaux_Cadoli_Mancini_AAAI_05} is an extension of a constraint satisfaction problem in which some of the variables are universally quantified (since the remaining variables are still existentially quantified).
In this latter framework, variables take value in discrete domains.
Universally quantified variables may be considered to represent certain kind of uncertainty: a choice of nature or an opponent.
A QCSP can formalize many AI problems including planning under uncertainty and playing a game against an opponent.
In this second application, the goal of the QCSP is to make a robust plan against the opponent.
Whereas finding a solution of a CSP is generally NP-complete, finding a solution for a QCSP is generally PSPACE-complete~\cite{Chen_PhD_04}.

Most of the recent decision procedure for QCSP~\cite{Bacchus_Stergiou_CP_07,Benedetti_Lallouet_Vautard_RAC_06,Gent_Nightingale_Stergiou_IJCAI_05,Bordeaux_Monfroy_CP_02} are based on a search algorithm (except~\cite{Verger_Bessiere_JFPC_06} which is based on a bottom-up approach and~\cite{Gent_Nightingale_Rowley_ECAI_04} which is based on a translation to quantified boolean formulas) and off-line procedures (except~\cite{Baba_Joe_Iwasaki_Yokoo_IJCAI_11} which is an on-line real-time algorithm based on Monte Carlo game tree search and~\cite{Stynes_Brown_CP_09} which is based on standard game tree search techniques).
Such an algorithm chooses a variable, branches on the different values of the domain, verifies if the subproblems have some solutions and combines, according to the semantics of the quantifier associated to the variable, those solutions into a solution to the problem.

Knowledge compilation is considered in many AI applications where quick on-line responses are expected. 
In general, a knowledge base is compiled off-line into a target language which is then used on-line to answer some queries.
The goal is to have a lesser complexity for the query computation of the compiled knowledge base than for the initial knowledge base.
This principle is for example applied in product configuration where the set of possible configurations is compiled~\cite{Amilhastre_Fargier_Marquis_AI_02}.

As far as we know, the problem of compiling a knowledge base represented as a QCSP has not been treated but only for the related domain of quantified Boolean formulas~\cite{Stephan_DaMota_ICLA_09,Fargier_Marquis_AAAI_06}.
Our first contribution is a new formalism as compilation target language: the QCSP base.
Our second contribution is a definition of an optimality property for QCSP bases in order to give a polytime answer to the next move choice problem~\cite{Stephan_DaMota_ICLA_09} which raises the issue of whether one can change for another solution during the game.
Our third contribution is a compilation algorithm embedded in a search algorithm which is proved to compile a QCSP in an optimal QCSP base.

This article is organized as follows: 
Section~\ref{sec:preliminaires} establishes the necessary preliminaries, 
section~\ref{sec:definitions} presents our framework and target language for the compilation of QCSP, 
section~\ref{sec:construction} specifies an algorithm to compile a QCSP in our target language, 
section~\ref{sec:conclusion} concludes with a discussion and some further works.

\section{Preliminaries}
\label{sec:preliminaires}
Symbol \ILEXISTE. stands for existential quantifier and symbol \QUELQUESOIT. stands for universal quantifier.
Symbol \ET. stands for logical conjunction, symbol \TOP. stands for what is always true and symbol \BOTTOM. stands for what is always false.
A QCSP is a tuple  \(\qcsp{\SETVARIABLES.}{\RANK.}{\QUANT.}{\SETDOMAINES.}{\SETCONSTRAINTS.}\):
\SETVARIABLES. is a set of \(n\) variables, 
\RANK. is a bijection from \SETVARIABLES. to \([1..n]\), 
\QUANT. is a mapping from \SETVARIABLES. to \(\{\ILEXISTE.,\QUELQUESOIT.\}\) 
(\(\quant{v}\) is the quantifier associated to the variable \(v\)), 
\SETDOMAINES. is a mapping from \SETVARIABLES. to a set of domains \(\{D(v_1),\dots,D(v_n)\}\) where, for every variable \(v_i\in \SETVARIABLES.\), \(D(v_i)\) is the finite domain of all the possible values
(\(D(v)\) is the domain associated to the variable \(v\)), 
\SETCONSTRAINTS. is a set of contraints.
If \(v_{j_1},\dots,v_{j_m}\) are the variables of a constraint \(c_j\in \SETCONTRAINTES.\) then the relation associated to \(c_j\) is a subset of the Cartesian product \(D(v_{j_1}) \times \dots \times D(v_{j_m})\).
In what follows, we denote for every \(i\in [1..n]\), \(\QUANTIFICATEUR._i= \quant{v_i}\) and \(D_i= D(v_i)\).
A QCSP \qcsp{\SETVARIABLES.}{\RANK.}{\QUANT.}{\SETDOMAINES.}{\SETCONSTRAINTS.} on \(n\) variables will be denoted as follows to simplify notation:
\[\QUANTIFICATEUR._1v_1\dots\QUANTIFICATEUR._nv_n\bigwedge_{c_j\in \SETCONTRAINTES.}c_j\]
with  \(v_1 \in {D_1}\), \dots, \(v_n \in {D_n}\),  \(\rank{v_i}=i\), for every \(i\in [1..n]\)~;
\(\QUANTIFICATEUR._1v_1\dots\QUANTIFICATEUR._nv_n\) is the binder.

The QCSP \(\qcsp{\{x,y,z,t\}}{\RANK.}{\QUANT.}{\{\{0,1,2\}\}}{\SETCONTRAINTES.}\) with 
\[\left\{
\begin{array}{l}
\RANK. = \{(1,x),(2,y),(3,z),(4,t)\}, \\
\QUANT. = \{(x,\ILEXISTE.),(y,\ILEXISTE.),(z,\QUELQUESOIT.),(t,\ILEXISTE.)\}, \\
D(x)=D(y)=D(z)=D(t)=\{0,1,2\} \mbox{ and }\\
\SETCONTRAINTES. = \{(x = (y*z)+t)\}
\end{array}
\right.\]
 is, for example, denoted~:
\(\ILEXISTE.x\ILEXISTE.y\QUELQUESOIT.z\ILEXISTE.t (x = (y*z)+t)\)
with \(x,y,z,t \in {\{0,1,2\}}\).

In a binder, a maximal homogeneous sequence of quantifiers forms a bloc~; the first one (and also the outermost) is the leftmost.

The set \(\mathcal{T}_{i}(\BINDER.)\) with \(v_1 \in {D_1}\), \dots, \(v_n \in {D_n}\), \(\BINDER.= \QUANTIFICATEUR._1v_1\dots\QUANTIFICATEUR._nv_n\) for \(1\leq i\leq n\) is the set of trees such that
\begin{itemize}
\item every leaf node is labeled by the symbol $\Box$ and is at depth $i$,
\item every internal node at depth $k$, $0\leq k <i-1$ is labeled with the variable $v_{k+1}$,
\item every edge linking a node at depth $k$ to one of its children's nodes is labeled with an element of \(D_k\),
\item all the labels of the edges linking a node to its children nodes are different.
\end{itemize}


The following tree is, for example, an element of the set \(\arb{2}{\ILEXISTE.x\ILEXISTE.y\QUELQUESOIT.z\ILEXISTE.t}\) with \(x,y,z,t \in {\{0,1,2\}}\)~:


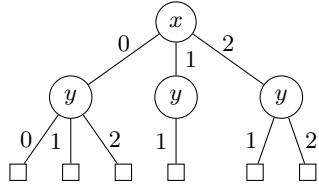
\begin{center}
{
\small
\tikzstyle{level 1}=[level distance=1cm, sibling distance=1.4cm]
\tikzstyle{level 2}=[level distance=1cm, sibling distance=0.7cm]
\begin{tikzpicture}[grow = down]
\node [circle,draw] (t0) {$x$}
child {node [circle,draw] (t0x0) {$y$}
  child {node [draw] (t0x0y0) {}
    edge from parent node[left] {0}
  }
  child {node [draw] (t0x0y1) {}
    edge from parent node[left] {1}
  }
  child {node [draw] (t0x0y2) {}
    edge from parent node[right] {2}
  }
  edge from parent node[above] {0}
}
child {node [circle,draw] (t0x1) {$y$}
  child {node [draw] (t0x1y1) {}
    edge from parent node[left] {1}
  }
  edge from parent node[right] {1}
}
child {node [circle,draw] (t0x2) {$y$}
  child {node [draw] (t0x2y1) {}
    edge from parent node[left] {1}
  }
  child {node [draw] (t0x2y1) {}
    edge from parent node[right] {2}
  }
  edge from parent node[above] {2}
};
\end{tikzpicture}
}
\end{center}

Let \(\qcsp{\SETVARIABLES.}{\RANK.}{\QUANT.}{\SETDOMAINES.}{\SETCONSTRAINTS.}\) be a QCSP such that \(\SETVARIABLES.=\{v_1,\dots,v_n\}\), with  \(v_1 \in {D_1}\), \dots, \(v_n \in {D_n}\), then a scenario is the sequence of the labels \(val_1,\dots,val_n\) on the path \((v_1,val_1),\dots,(v_n,val_n)\), \(val_i\in D_i\) for every \(i\), \(1\leq i\leq n\), of a tree of \(\arb{n}{\QUANTIFICATEUR._1v_1\dots\QUANTIFICATEUR._nv_n}\) and a strategy is a tree of \(\arb{n}{\QUANTIFICATEUR._1v_1\dots\QUANTIFICATEUR._nv_n}\) such that
\begin{itemize}
\item every node labeled with an existentially quantified variable has a unique child node and  
\item every node labeled with a universally quantified variable whose associated domain is of size $k$ admits $k$ children nodes.
\end{itemize}
A scenario \(val_1,\dots,val_n\) for a QCSP \(\qcsp{\SETVARIABLES.}{\RANK.}{\QUANT.}{\SETDOMAINES.}{\SETCONSTRAINTS.}\) such that \(\SETVARIABLES.=\{v_1,\dots,v_n\}\) is a winning scenario if \((\bigwedge_{1\leq i \leq n}v_i=val_i) \ET. (\bigwedge_{c_j\in \SETCONTRAINTES.}c_j)\) is true~; such a scenario corresponds to the complete instantiation \([v_1\leftarrow val_1]\), \dots, \([v_n\leftarrow val_n]\) ; it is a winning scenario if the instantiation satisfies all the constraints.
A strategy is a winning strategy if all the scenarios are winning scenarios.
If there is no quantifier, the $\Box$ strategy is always a winning strategy.

The scenario \(0,0,2,0\), which corresponds to the complete instantiation \([x\leftarrow 0]\), \([y\leftarrow 0]\), \([z\leftarrow 2]\) and \([t\leftarrow 0]\), is a winning scenario, since \(0 = (0*2)+0\).
The following strategy is a winning strategy 

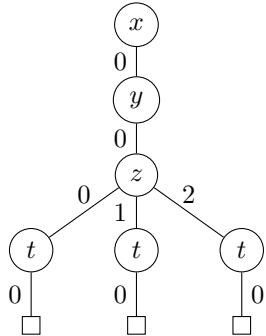
\begin{center}
{
\tikzstyle{level 1}=[level distance=1cm, sibling distance=1cm]
\tikzstyle{level 2}=[level distance=1cm, sibling distance=1cm]
\tikzstyle{level 3}=[level distance=1cm, sibling distance=1.4cm]
\tikzstyle{level 4}=[level distance=1cm, sibling distance=1cm]

\begin{tikzpicture}[grow = down]
\node [circle,draw] (root){$x$} 
child {node [circle,draw] (x0) {$y$}
  child {node [circle,draw] (x0y0) {$z$}
    child {node [circle,draw] (x0y0z2) {$t$}
      child {node [draw] (x0y0z2t0) {} edge from parent node[left] {0}}
      edge from parent node[above] {0}
    }
    child {node [circle,draw] (x0y0z1) {$t$}
      child {node [draw] (x0y0z1t0) {} edge from parent node[left] {0}}
      edge from parent node[left] {1}
    }   
    child {node [circle,draw] (x0y0z0) {$t$}
      child {node [draw] (x0y0z0t0) {} edge from parent node[right] {0}}
      edge from parent node[above] {2}
    }
    edge from parent node[left] {0}
  }
  edge from parent node[left] {0}
};
\end{tikzpicture}
}
\end{center}

for the QCSP
\(\ILEXISTE.x\ILEXISTE.y\QUELQUESOIT.z\ILEXISTE.t (x = (y*z)+t), x,y,z,t\in {\{0,1,2\}}\)
since \((0 = (0*0)+0)\), \((0 = (0*1)+0)\) and \((0 = (0*2)+0)\).

We can give a more intuitive and recursive decision semantics for QCSP as follows: A QCSP \(\QUELQUESOIT. x\BINDER. C\) with \(x\in D\) admits a winning strategy if and only if, for every \(val \in D\), \(\BINDER. (C\ET. (x=val))\) admits a winning strategy and a QCSP \(\ILEXISTE. x\BINDER. C\) with \(x\in D\) admits a winning strategy if and only if, for at least one \(val \in D\), \(\BINDER. (C\ET. (x=val))\) admits a winning strategy.

\section{Base for QCSP}
\label{sec:definitions}
From a complexity point of view and under some classical assumptions,  winning strategies are exponential in space in worst case w.r.t. the number of variables of the QCSP~\cite{Coste_LeBerre_Letombe_Marquis_JSAT_06}. 
But the number of winning strategies may also be exponential in worst case.
A naive way to compile a QCSP would be to store all the winning strategies in a set but this approach is intractable in practice.
For example, the  QCSP \(\QUELQUESOIT.x\QUELQUESOIT.y\ILEXISTE.z\ILEXISTE.t (x = (y*z)+t), x,y,z,t\in {\{0,1,2\}}\) admits 324 winning strategies.
Another way is to store a tree which contains only the scenarios present in the winning strategies.
This approach is not very useful too from the knowledge representation point of view since there is no direct access to the possibilities of an existentially quantified variable except for those of the first bloc.

We define in this section our formalism as a target language for QCSP compilation: the {\it QCSP base}.
We also define the semantics of QCSP bases in terms of QCSP.
We introduce a property of optimality for QCSP bases and prove a very interesting result about optimal QCSP.

\subsection{Definitions for QCSP bases}
Intuitively, a QCSP base is a set of strategies organized according to a mechanism of guards for every existentially quantified variable and every value of the domain.
Such a guard is a pair of a value and a tree which is the expression of what have already been played by both opponents.

\begin{definition}[QCSP base]
\label{def:base}
A QCSP base is either
\begin{itemize}
\item the symbol \BLTOP.
\item the symbol \BLBOTTOM.
\item a pair \(\bl{\BINDER.}{G}\) with  \(n > 0\), \(\BINDER.= \QUANTIFICATEUR._1v_1\dots\QUANTIFICATEUR._nv_n\) and \(G = [G_{e_1} ,\dots, G_{e_m}]\) a list such that 
\begin{itemize} 
\item \(e_1, \dots, e_m\) is the set of indexes of the existentially quantified  variables\footnote{\(u_1, \dots, u_p\) is the set of indexes of the universally quantified  variables, \(\{e_1, \dots, e_m\}\cup \{u_1, \dots, u_p\} = [1..n]\), \(\{e_1, \dots, e_m\}\cap \{u_1, \dots, u_p\} = \emptyset\), \(\quant{v_{e_i}}= \ILEXISTE.\), for every \(i\), \(1\leq i \leq n\), \(\quant{v_{u_i}}= \QUELQUESOIT.\), for every \(i\), \(1 \leq i\leq p\).}~;
\item every \(G_{e_k}\), \(1 \leq k\leq m\) is a function with non-empty graph \(\{\fonc{val_1}{T_1},\dots, \fonc{val_{j_k}}{T_{j_k}}\}\), \(val_1,\dots, val_{j_k}\in D(v_{e_k})\) and \(T_1,\dots,T_{j_k} \in \mathcal{T}_{e_k}(\BINDER.)\).
\end{itemize}
A pair \((val,T) \in G_{e_k}\) is a guard for the existentially quantified variable \(v_{e_k}\).
\end{itemize}

\end{definition}

In what follows, the QCSP bases \BLTOP. and \BLBOTTOM. are semantically interpreted as respectively what is always true and what is always false and algorithmically as respectively what admits every strategy as a winning strategy and what admits no winning strategy at all.

\begin{example}
\label{exe:base_EEAE}


The following guard sets \(G_x\), \(G_y\) and \(G_z\) are guard sets of the QCSP base \(B=\bl{\ILEXISTE.x\ILEXISTE.y\QUELQUESOIT.z\ILEXISTE.t}{[G_x,G_y,G_t]}\) with \(x,y,z,t\in {\{0,1,2\}}\).

\(\begin{array}{ll}
\\
G_x = & [(0,\Box),(1,\Box),(2,\Box)],
\\
\\
G_y = & [(0,
{
\tikzstyle{level 1}=[level distance=1cm, sibling distance=0.7cm]
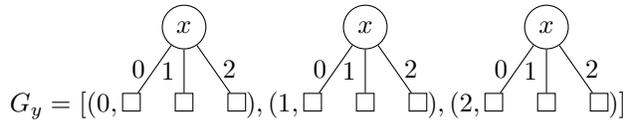
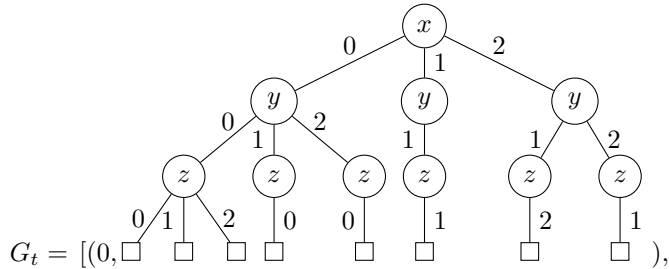
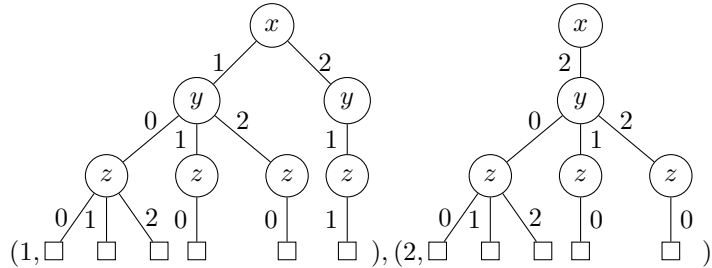
\begin{tikzpicture}[grow = down]
\node [circle,draw] (root) {$x$}
child {node [draw] (x0y2) {}
  edge from parent node[left] {0}
}
child {node [draw] (x0y1) {}
  edge from parent node[left] {1}
}
child {node [draw] (x0y0) {}
  edge from parent node[right] {2}
};
\end{tikzpicture}
}
),(1,
{
\tikzstyle{level 1}=[level distance=1cm, sibling distance=0.7cm]
\begin{tikzpicture}[grow = down]
\node [circle,draw] (root) {$x$}
child {node [draw] (x0y2) {}
  edge from parent node[left] {0}
}
child {node [draw] (x0y1) {}
  edge from parent node[left] {1}
}
child {node [draw] (x0y0) {}
  edge from parent node[right] {2}
};
\end{tikzpicture}
}
),(2,
{
\tikzstyle{level 1}=[level distance=1cm, sibling distance=0.7cm]
\begin{tikzpicture}[grow = down]
\node [circle,draw] (root) {$x$}
child {node [draw] (x0y2) {}
  edge from parent node[left] {0}
}
child {node [draw] (x0y1) {}
  edge from parent node[left] {1}
}
child {node [draw] (x0y0) {}
  edge from parent node[right] {2}
};
\end{tikzpicture}
}
)]
\\
\\
G_t = &

[(0,
{
\tikzstyle{level 1}=[level distance=1.0cm, sibling distance=2cm]
\tikzstyle{level 2}=[level distance=1.0cm, sibling distance=1.2cm]
\tikzstyle{level 3}=[level distance=1.0cm, sibling distance=0.7cm]
\begin{tikzpicture}[grow = down]
\node [circle,draw] (t0) {$x$}
child {node [circle,draw] (t0x0) {$y$}
  child {node [circle,draw] (t0x0y0) {$z$}
    child {node [draw] (t0x0y0z0) {} edge from parent node[left] {0}}
    child {node [draw] (t0x0y0z1) {} edge from parent node[left] {1}}
    child {node [draw] (t0x0y0z2) {} edge from parent node[right] {2}}
    edge from parent node[above] {0}
  }
  child {node [circle,draw] (t0x0y1) {$z$}
    child {node [draw] (t0x0y0z2) {} edge from parent node[right] {0}}
    edge from parent node[left] {1}
  }
  child {node [circle,draw] (t0x0y2) {$z$}
    child {node [draw] (t0x0y0z0) {} edge from parent node[left] {0}}
    edge from parent node[above] {2}
  }
  edge from parent node[above] {0}
}
child {node [circle,draw] (t0x1) {$y$}
  child {node [circle,draw] (t0x1y1) {$z$}
    child {node [draw] (t0x2y1z1) {} edge from parent node[right] {1}}
    edge from parent node[left] {1}
  }
  edge from parent node[right] {1}
}
child {node [circle,draw] (t0x2) {$y$}
  child {node [circle,draw] (t0x2y1) {$z$}
    child {node [draw] (t0x2y1z2) {} edge from parent node[right] {2}}
    edge from parent node[left] {1}
  }
  child {node [circle,draw] (t0x2y1) {$z$}
    child {node [draw] (t0x2y12z1) {} edge from parent node[right] {1}}
    edge from parent node[right] {2}
  }
  edge from parent node[above] {2}
};
\end{tikzpicture}
}
),\\ \\ &(1,
{
\tikzstyle{level 1}=[level distance=1.0cm, sibling distance=2cm]
\tikzstyle{level 2}=[level distance=1.0cm, sibling distance=1.2cm]
\tikzstyle{level 3}=[level distance=1.0cm, sibling distance=0.7cm]

\begin{tikzpicture}[grow = down]
\node [circle,draw] (t1) {$x$}
child {node [circle,draw] (t1x2) {$y$}
  child {node [circle,draw] (t1x2y1) {$z$}
    child {node [draw] (t1x1y0z0) {} edge from parent node[left] {0}}
    child {node [draw] (t1x1y0z1) {} edge from parent node[left] {1}}
    child {node [draw] (t1x2y1z1) {} edge from parent node[right] {2}}
    edge from parent node[above] {0}
  }
  child {node [circle,draw] (t1x2y1) {$z$}
    child {node [draw] (t1x1y0z0) {} edge from parent node[left] {0}}
    edge from parent node[left] {1}
  }
  child {node [circle,draw] (t1x2y1) {$z$}
    child {node [draw] (t1x1y0z0) {} edge from parent node[left] {0}}
    edge from parent node[above] {2}
  }
  edge from parent node[left] {1}
}
child {node [circle,draw] (t1x1) {$y$}
  child {node [circle,draw] (t1x1y0) {$z$}
    child {node [draw] (t1x1y0z0) {} edge from parent node[left] {1}}
    edge from parent node[left] {1}
  }
  edge from parent node[right] {2}
};
\end{tikzpicture}
}
), (2,

{

\tikzstyle{level 1}=[level distance=1.0cm, sibling distance=2cm]
\tikzstyle{level 2}=[level distance=1.0cm, sibling distance=1.2cm]
\tikzstyle{level 3}=[level distance=1.0cm, sibling distance=0.7cm]
\begin{tikzpicture}[grow = down]
\node [circle,draw] (t2) {$x$}
child {node [circle,draw] (t2x2) {$y$}
  child {node [circle,draw] (t2x2y0) {$z$}
    child {node [draw] (t2x2y0z0) {} edge from parent node[left] {0}}
    child {node [draw] (t2x2y0z1) {} edge from parent node[left] {1}}
    child {node [draw] (t2x2y0z2) {} edge from parent node[right] {2}}
    edge from parent node[above] {0}
  }
  child {node [circle,draw] (t2x2y1) {$z$}
    child {node [draw] (t2x2y1z0) {} edge from parent node[right] {0}}
    edge from parent node[right] {1}
  }
  child {node [circle,draw] (t2x2y1) {$z$}
    child {node [draw] (t2x2y1z0) {} edge from parent node[right] {0}}
    edge from parent node[above] {2}
  }
  edge from parent node[left] {2}
};
\end{tikzpicture}
}
)
\end{array}
\)

\end{example}

\subsection{Interpretation}
The semantics of a QCSP base is expressed by an interpretation to the QCSP.
First of all, we interpret the trees of the guards of the QCSP bases as tuples of values.

\begin{definition}[interpretation of a tree]
\label{def:inter_arbre}
The interpretation of a tree $T$ of a guard \((val,T)\) according to a value $val$ is the set 
\[\Intb{val}{T}= \{(val, e_1, \dots, e_n) | e_1\dots e_n \mbox{ a branch of a tree }T\}.\]
In particular, \(\Intb{val}{\Box}= \{val\}\).
The interpretation of a set $G$ of guards (value, tree) is by extension~:
\[\Intu{G} = \bigcup_{(val,T) \in G}\Intb{val}{T}.\]

\end{definition}

\begin{example}
\label{exe:inter_arbre}

{\em (Example~\ref{exe:base_EEAE} continued.)} The interpretation of the tree $T$ extracted from the set of guards \(G_t\)~: 


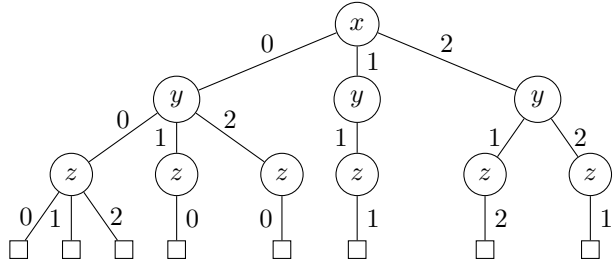
\tikzstyle{level 1}=[level distance=1cm, sibling distance=2.4cm]
\tikzstyle{level 2}=[level distance=1cm, sibling distance=1.4cm]
\tikzstyle{level 3}=[level distance=1cm, sibling distance=0.7cm]

\begin{center}
{
\begin{tikzpicture}[grow = down]
\node [circle,draw] (t0) {$x$}
child {node [circle,draw] (t0x0) {$y$}
  child {node [circle,draw] (t0x0y0) {$z$}
    child {node [draw] (t0x0y0z0) {} edge from parent node[left] {0}}
    child {node [draw] (t0x0y0z1) {} edge from parent node[left] {1}}
    child {node [draw] (t0x0y0z2) {} edge from parent node[right] {2}}
    edge from parent node[above] {0}
  }
  child {node [circle,draw] (t0x0y1) {$z$}
    child {node [draw] (t0x0y0z2) {} edge from parent node[right] {0}}
    edge from parent node[left] {1}
  }
  child {node [circle,draw] (t0x0y2) {$z$}
    child {node [draw] (t0x0y0z0) {} edge from parent node[left] {0}}
    edge from parent node[above] {2}
  }
  edge from parent node[above] {0}
}
child {node [circle,draw] (t0x1) {$y$}
  child {node [circle,draw] (t0x1y1) {$z$}
    child {node [draw] (t0x2y1z1) {} edge from parent node[right] {1}}
    edge from parent node[left] {1}
  }
  edge from parent node[right] {1}
}
child {node [circle,draw] (t0x2) {$y$}
  child {node [circle,draw] (t0x2y1) {$z$}
    child {node [draw] (t0x2y1z2) {} edge from parent node[right] {2}}
    edge from parent node[left] {1}
  }
  child {node [circle,draw] (t0x2y1) {$z$}
    child {node [draw] (t0x2y12z1) {} edge from parent node[right] {1}}
    edge from parent node[right] {2}
  }
  edge from parent node[above] {2}
};
\end{tikzpicture}
}
\end{center}

according to the value 0 is the set 

\[\begin{array}{ll}
\lefteqn{\Intb{0}{T} = \{(0,0,0,0),(0,0,0,1),(0,0,0,2),}\\
& (0,0,1,0),(0,0,2,0),(0,1,1,1),\\
& (0,2,1,2),(0,2,2,1)\}.
\end{array}\]

and

\[\begin{array}{ll}
\lefteqn{\Intu{G_t} = \{(0,0,0,0),(0,0,0,1),(0,0,0,2),}\\
&(0,0,1,0),(0,0,2,0),(0,1,1,1),(0,2,1,2),\\
&(0,2,2,1),(1,1,0,0),(1,1,0,1),(1,1,0,2),\\
&(1,1,1,0),(1,1,2,0),(1,2,1,1),(2,2,0,0),\\
&(2,2,0,1),(2,2,0,2),(2,2,1,0),(2,2,2,0)\}
\end{array}
\]

One can remark that for all \((val_t,val_x,val_y,val_z)\in \Intu{G_t}\), the instantiation
\([t \leftarrow val_t][x \leftarrow val_x][y\leftarrow val_y][z \leftarrow val_z]\) satisfies the constraint \((x = (y*z)+t)\).

\end{example}

We now define the interpretation of a QCSP base.

\begin{definition}[interpretation of a QCSP base]
\label{def:inter_base}
The interpretation function \((\cdot)^*\) of a QCSP base to a QCSP is defined as follows (\(\BINDER.= \QUANTIFICATEUR._1v_1\dots\QUANTIFICATEUR._nv_n\))~:

\(
\begin{array}{ll}
\lefteqn{(\BLTOP.)^*= \TOP.}\\
\lefteqn{(\BLBOTTOM.)^* = \BOTTOM.}\\
\lefteqn{(\bl{\BINDER.}{[G_{e_1} ,\dots, G_{e_m}]})^*=}\\
& \BINDER.\bigwedge_{e_i\in [e_1, \ldots, e_m]}((v_{e_k},v_1,\dots,v_{e_k-1})\in \Intu{G_{e_k}})
\end{array}
\)

\end{definition}

The interpretation of a QCSP base is a QCSP but only on table constraints.

\begin{example}
\label{exe:inter_base}
{\em (Examples~\ref{exe:base_EEAE} and~\ref{exe:inter_arbre} continued.)}

\[
(B)^*= \ILEXISTE.x\ILEXISTE.y\QUELQUESOIT.z\ILEXISTE.t((x\in \Intu{G_x}) \ET. ((y,x)\in \Intu{G_y}) \ET. ((t,x,y,z)\in \Intu{G_t}))
\]

with \(x,y,z,t\in {\{0,1,2\}}\), \(\Intu{G_x}=\{0, 1, 2\}\) and \(\Intu{G_y} = \{0,1,2\}^2\).

\end{example}

\subsection{Properties of QCSP bases}
To a given QCSP, many different QCSP bases may be such that their interpretations have exactly the same set of winning strategies as that QCSP.
We define this property as the {\it compatibility property}.

\begin{definition}[compatibility of a QCSP base]
\label{def:compatibilite}
A QCSP base is compatible with a QCSP if its interpretation has exactly the same winning strategy.

\end{definition}

In what follows, we will see that the set of compatible QCSP bases may be seen as a good candidate as a target for a compilation language.

\begin{example}
\label{exe:compatibilite}
{\em (Examples~\ref{exe:base_EEAE},~\ref{exe:inter_arbre} and~\ref{exe:inter_base} continued.)}
The QCSP base \(B\) is compatible with the QCSP \(\ILEXISTE.x\ILEXISTE.y\QUELQUESOIT.z\ILEXISTE.t (x = (y*z)+t)\) with \(x,y,z,t \in {\{0,1,2\}}\).
If, for example, the pair \((2,\Box)\) of \(G_x\) is discarded then the resulting QCSP base is no more compatible with the QCSP since two of the four winning strategies are lost.

\end{example}

The following theorem establishes immediately the completeness of the QCSP base formalism w.r.t. QCSP. 

\begin{theorem}[completeness]
\label{th:compatibilite}
For every QCSP there exists a compatible base.

\end{theorem}



When a QCSP represents a finite two-player game, one of the most important issues for the existential player, at each turn during the game, is the following:
``What do I have to play to be certain to win the game?''
If a winning strategy has been already computed before the game begins, the player has only to follow it.
But if the uncertainty was not completely known and if the current winning strategy can not be applied anymore, the existential player has to compute again a new strategy and has to pay also the complete algorithmic price.

\begin{definition}[next move choice problem]
\label{def:prochain_mouvement}
The {\em next move choice problem} is defined as follows.~\\
\begin{itemize}
\item {\it Instance}~: A QCSP 
\(\QUANTIFICATEUR._1 v_1\dots\QUANTIFICATEUR._nv_n\bigwedge_{c\in \SETCONTRAINTES.}c\)
with \(v_1 \in {D_1}\), \dots, \(v_n \in {D_n}\) and a sequence of instantiations \(\substxparysq{v_1}{val_1}, \dots, \substxparysq{v_i}{val_i}\) obtained from a winning strategy  for a QCSP with \(\quant{v_i} = \ILEXISTE.\) and \(val_1\in D_1\), \dots, \(val_i \in D_i\).
\item {\it Query}~: Is there any winning strategy for a QCSP
\[
\QUANTIFICATEUR._{i+1} v_{i+1}\dots\QUANTIFICATEUR._nv_n
\bigwedge_{c\in \SETCONTRAINTES.}c\ET. (v_1=val_1) \ET. (v_{i-1}=val_{i-1})\ET. (v_i=val_i')
\]
with \(v_{i+1} \in {D_{i+1}}\), \dots, \(v_n \in {D_n}\), \(val_i'\in D_i\), \(val_i'\neq val_i\)~?
\end{itemize}

\end{definition}

Clearly enough the next move choice problem is still a PSPACE-complete problem
since \(\QUANTIFICATEUR._{i+1} v_{i+1}\dots\QUANTIFICATEUR._nv_n\bigwedge_{c\in \SETCONTRAINTES.}c\ET. (v_1=val_1) \ET. (v_{i-1}=val_{i-1})\ET. (v_i=val_i')\) with \(v_{i+1} \in {D_{i+1}}\), \dots, \(v_n \in {D_n}\), \(val_i'\in D_i\), \(val_i'\neq val_i\) is a QCSP.

We introduce a new property for a QCSP base which guarantees that the next move choice problem is no more PSPACE-complete but polytime w.r.t. the size of the QCSP base.
A QCSP base is {\em optimal} if all the guards associated to the moves played by the existential player are verified then this player is sure to follow a winning strategy.

\begin{definition}[optimality]
\label{def:optimalite}
Let \(B = \bl{\QUANTIFICATEUR._1v_1\dots\QUANTIFICATEUR._nv_n}{[G_{e_1} ,\dots, G_{e_m}]}\) be a QCSP base and \((B)^*= \QUANTIFICATEUR._1v_1\dots\QUANTIFICATEUR._nv_n C\) with \(v_1 \in {D_1}\), \dots, \(v_n \in {D_n}\).
This base is optimal if the following property is verified.
For every \(i\), \(i\in [1 \dots m]\), let \(C_i\) be the set of constraints \(\{(v_{e_k} = val_{e_k}) | 1\leq k < i\}\) such that \((val_{e_k},val_{e_1}, \dots, val_{e_{k-1}}) \in \Intb{val_{e_k}}{a_{e_{k}}}\), \((val_{e_{k}},a_{e_{k}}) \in G_{e_{k}}\).

Then for every guard \((val, a) \in G_{e_i}\), \((val,val_{e_1}, \dots, val_{e_{i-1}}) \in \Intb{val}{a}\) if and only if \(\QUANTIFICATEUR._{e_i+1}v_{e_i+1}\dots\QUANTIFICATEUR._nv_n (C \ET. (v_{e_i}=val) \ET. \bigwedge_{c\in C_i}c)\) admits a winning strategy.

\end{definition}

The underlying order of this notion of optimality is the number of winning scenarios which are not a branch of any winning strategy.
In case of the interpretation of an optimal base this number is zero.

\begin{example}
\label{exe:base_EEAE_opt}
The following guard sets \(G^{opt}_x\), \(G^{opt}_y\) and \(G^{opt}_t\) are guard sets for the QCSP base \(B^{opt}=\bl{\ILEXISTE.x\ILEXISTE.y\QUELQUESOIT.z\ILEXISTE.t}{[G^{opt}_x,G^{opt}_y,G^{opt}_t]}\) which is optimal and compatible with the QCSP~:  
\(\ILEXISTE.x\ILEXISTE.y\QUELQUESOIT.z\ILEXISTE.t (x = (y*z)+t)\)
 with \(x,y,z,t\in {\{0,1,2\}}\).

\begin{figure}

\(\begin{array}{l}
G^{opt}_x = [(0,\Box),(1,\Box),(2,\Box)]\\


\\
\\
\tikzstyle{level 1}=[level distance=1cm, sibling distance=0.7cm]

G^{opt}_y = [(0, T_0^{y}=
{
\begin{tikzpicture}[grow = down]
\node [circle,draw] (root) {$x$}
child {node [draw] (x0y2) {}
  edge from parent node[left] {0}
}
child {node [draw] (x0y1) {}
  edge from parent node[left] {1}
}
child {node [draw] (x0y0) {}
  edge from parent node[right] {2}
};
\end{tikzpicture}
}
),(1, T_1^{y}=
{
\begin{tikzpicture}[grow = down]
\node [circle,draw] (root) {$x$}
child {node [draw] (x1y1) {} 
  edge from parent node[left] {2}
};
\end{tikzpicture}
}
)]

\\
\\
G^{opt}_t= [(0, T_0^{t}=
{

\tikzstyle{level 1}=[level distance=1.0cm, sibling distance=1.6cm]
\tikzstyle{level 2}=[level distance=1.0cm, sibling distance=1.0cm]
\tikzstyle{level 3}=[level distance=1.0cm, sibling distance=0.7cm]

\begin{tikzpicture}[grow = down]
\node [circle,draw] (t0) {$x$}
child {node [circle,draw] (t0x0) {$y$}
  child {node [circle,draw] (t0x0y0) {$z$}
    child {node [draw] (t0x0y0z0) {} edge from parent node[left] {0}}
    child {node [draw] (t0x0y0z1) {} edge from parent node[left] {1}}
    child {node [draw] (t0x0y0z2) {} edge from parent node[right] {2}}
    edge from parent node[left] {0}
  }
  edge from parent node[left] {0}
}
child {node [circle,draw] (t0x2) {$y$}
  child {node [circle,draw] (t0x2y1) {$z$}
    child {node [draw] (t0x2y1z2) {} edge from parent node[right] {2}}
    edge from parent node[left] {1}
  }
  edge from parent node[right] {2}
};
\end{tikzpicture}
}
),\\(1, T_1^{t}=

{

\tikzstyle{level 1}=[level distance=1.0cm, sibling distance=1.6cm]
\tikzstyle{level 2}=[level distance=1.0cm, sibling distance=1.0cm]
\tikzstyle{level 3}=[level distance=1.0cm, sibling distance=0.7cm]
\begin{tikzpicture}[grow = down]
\node [circle,draw] (t1) {$x$}
child {node [circle,draw] (t1x1) {$y$}
  child {node [circle,draw] (t1x1y0) {$z$}
    child {node [draw] (t1x1y0z0) {} edge from parent node[left] {0}}
    child {node [draw] (t1x1y0z1) {} edge from parent node[left] {1}}
    child {node [draw] (t1x1y0z2) {} edge from parent node[right] {2}}
    edge from parent node[left] {0}
  }
  edge from parent node[left] {1}
}
child {node [circle,draw] (t1x2) {$y$}
  child {node [circle,draw] (t1x2y1) {$z$}
    child {node [draw] (t1x2y1z1) {} edge from parent node[right] {1}}
    edge from parent node[left] {1}
  }
  edge from parent node[right] {2}
};
\end{tikzpicture}
}
),(2, T_2^{t}=
{

\tikzstyle{level 1}=[level distance=1.0cm, sibling distance=1.6cm]
\tikzstyle{level 2}=[level distance=1.0cm, sibling distance=1.6cm]
\tikzstyle{level 3}=[level distance=1.0cm, sibling distance=0.7cm]

\begin{tikzpicture}[grow = down]
\node [circle,draw] (t2) {$x$}
child {node [circle,draw] (t2x2) {$y$}
  child {node [circle,draw] (t2x2y0) {$z$}
    child {node [draw] (t2x2y0z0) {} edge from parent node[left] {0}}
    child {node [draw] (t2x2y0z1) {} edge from parent node[left] {1}}
    child {node [draw] (t2x2y0z2) {} edge from parent node[right] {2}}
    edge from parent node[left] {0}
  }
  child {node [circle,draw] (t2x2y1) {$z$}
    child {node [draw] (t2x2y1z0) {} edge from parent node[right] {0}}
    edge from parent node[right] {1}
  }
  edge from parent node[left] {2}
};
\end{tikzpicture}
}
)]
\end{array}\)
\caption{Guard sets for an optimal QCSP base.\label{fig:base_EEAE_opt}}
\end{figure}

We explicit hereafter the optimality of the QCSP base but we use the QCSP constraint \((x = (y*z)+t)\) instead of the table constraints in order to simplify.
\begin{itemize}
\item \(i=1\) (i.e. \(v_{e_i}=x\)) then \(C_i=\emptyset\) and for every \(K\in \{0,1,2\}\), \(K \in \Intb{K}{\Box}=\{K\}\) and \(\ILEXISTE.y\QUELQUESOIT.z\ILEXISTE.t (x = (y*z)+t) \ET. (x = K)\),  with \(y,z,t\in {\{0,1,2\}}\), admits a winning strategy.
\item \(i=2\) (i.e. \(v_{e_i}=y\)) then
 \begin{itemize}
 \item for every \(K\in\{0,1,2\}\) \((K, \Box) \in G^{opt}_x\), \(\Intb{0}{T_0^y}= \{(0,0),(0,1),(0,2)\}\) and for every \(K\in\{0,1,2\}\), \((0,K)\in \Intb{0}{T_0^y}\) and  \(\QUELQUESOIT.z\ILEXISTE.t (x = (y*z)+t) \ET. (y=0) \ET. (x = K) \),  with \(z,t\in {\{0,1,2\}}\), admits a winning strategy~;
 \item \((1, \Box) \in G^{opt}_x\), \(\Intb{1}{T_1^y}= \{(1,1)\}\) and \((1,1)\in \Intb{1}{T_1^y}\) and  \(\QUELQUESOIT.z\ILEXISTE.t (x = (y*z)+t) \ET. (y=1) \ET. (x = 1) \),  with \(z,t\in {\{0,1,2\}}\), admits winning strategy~; 
\((1,0)\not\in \Intb{1}{T_1^y}\) and \(\QUELQUESOIT.z\ILEXISTE.t (x = (y*z)+t) \ET. (y=1) \ET. (x = 0)\),  with \(z,t\in {\{0,1,2\}}\), does not admit a winning strategy~; 
\((1,2)\not\in \Intb{1}{T_1^y}\) and \(\QUELQUESOIT.z\ILEXISTE.t (x = (y*z)+t) \ET. (y=1) \ET. (x = 2)\),  with \(z,t\in {\{0,1,2\}}\), does not admit a winning strategy~;
 \item for \(y=2\), the is no pair \((2,T^y_2)\in G_y\) and \(\ILEXISTE.x\QUELQUESOIT.z\ILEXISTE.t (x = (y*z)+t) \ET. (y=2)\), with \(x,z,t\in {\{0,1,2\}}\), does not admit a winning strategy.
 \end{itemize}
\item \(i=3\) (i.e. \(v_{e_i}=t\)) then (we only treat the case \(t=0\), the others are similar) 
 \begin{itemize}
 \item \((2, \Box) \in G^{opt}_x\), \((1, T^y_1)\in G^{opt}_y\) with \((1,2)\in \Intb{1}{T^y_1}\) and \((0,2,1,2)\in \Intb{0}{T^t_0}\) and \((x = (y*z)+t) \ET. (x=2)\ET. (y=1)\ET. (z=2)\ET. (t=0)\) admits a winning strategy~; it is similar for \((0,0,0,0)\), \((0,0,0,1)\) and \((0,0,0,2)\)~;
 \item for all the other cases \((0,val_x,val_y,val_z)\not\in \Intb{0}{T^t_0}\) and \((x = (y*z)+t) \ET. (x=val_x) \ET. (y = val_y) \ET. (z=val_z) \ET. (t=0)\) does not admit a winning strategy.
 \end{itemize}
\end{itemize}

\end{example}

The most important property of optimal QCSP base is that the next move choice problem for the interpretation of compatible optimal base with a QCSP is no more PSPACE-complete but polytime.

\begin{theorem}[next move choice problem]
\label{th:prochain_mouvement}
The next move decision problem for the interpretation of an optimal base is polytime in the size of the base.

\end{theorem}

\section{Compilation of a QCSP to an optimal QCSP base}
\label{sec:construction}
We present in this section an algorithm based on a search algorithm and establish that the result of the application of this algorithm is an optimal QCSP base compatible with the initial QCSP.
Algorithm~1 \RCQCSP. computes a compatible QCSP base from a QCSP following the inductive definition of the semantics of the QCSP.
This algorithm first computes a fix-point for the set of constraints and returns \BLBOTTOM. if a contradiction is detected.
If it is not the case and the binder is not empty then \BLTOP. is returned.
Otherwise for every value \(val\) of the domain of the outermost variable \(x\) of the binder, the constraint  \((x=val)\) is added to the constraint store and the algorithm is recursively called.
If the variable is universally quantified and at least one subproblem returns \BLBOTTOM. then \BLBOTTOM. is returned.
If the variable is existentially quantified and all the subproblems return \BLBOTTOM. then \BLBOTTOM. is returned.
In any other cases, operators \COMBILEXISTE. or \COMBQUELQUESOIT. are called to combine the resulting QCSP bases together.

\begin{myalgorithm}[h!]
\label{algo:rec_comp}
\caption{\RCQCSP.}
\begin{mycode}
\ENTREE{\(\BINDER. :\) a binder of a QCSP}
\ENTREE{\(C :\) a set of constraints of a QCSP}
\ENSURE{a QCSP base or \BLTOP. or \BLBOTTOM.}
\IF{\(reach\_fixpoint(C) = failure\)}
\RETURN{\(\BLBOTTOM.\)}
\ENDIF
\STATE{\textbf{if }\(\vide{\BINDER.}\) \textbf{ then return }\BLTOP. \textbf{ end if}}
\STATE{\(\QUANTIFICATEUR. x_{D} \leftarrow \head{\BINDER.}; {listValBase}\leftarrow{[]} ; d \leftarrow D\)}
\WHILE{\(! \vide{d}\)}
 \STATE{\(val \leftarrow \head{d} ; d \leftarrow \tail{d}\)}
 \STATE{\({base}\leftarrow {\rcqcsp{\tail{\BINDER.}}{C \cup\{x = val\}}}\)}
 \IF{\(base = \BLBOTTOM. \& \QUANTIFICATEUR. = \QUELQUESOIT.\)}
  \RETURN{\(\BLBOTTOM.\)}
 \ENDIF
 \STATE{\({listValBase}\leftarrow {[(val,base)|listValBase]}\)}
\ENDWHILE
 \IF{\(\vide{listValBase}\)}
 \RETURN{\BLBOTTOM.}
 \ENDIF
\IF{\(\QUANTIFICATEUR. = \ILEXISTE.\)}
 \RETURN{\({\combilexiste{x_{D}}{\BINDER.}{listValBase}}\)}
\ELSE
 \RETURN{\({\combquelquesoit{x_{D}}{\BINDER.}{listValBase}}\)}
\ENDIF
\end{mycode}
\end{myalgorithm}

Operators \COMBQUELQUESOIT. and \COMBILEXISTE. specified respectively by the algorithms~2 and~3 work as follows.
First we describe the \COMBQUELQUESOIT. operator. 
Function \(constants(l)\) checks if the list of pairs as argument \(l\) does not contain only \BLTOP. or \BLBOTTOM. for second element.
If the check \(constants(l)\) is verified, it is necessarily only a list of \BLTOP. associated with all the values of the domain of the variable in case of an innermost bloc of universal quantifiers and then \BLTOP. is returned.
Otherwise, the operator \COMBINE. defined hereafter is applied and the result is returned since the universally quantified variables are not associated to guards.
Now we describe the \COMBILEXISTE. operator. 
If the check \(constants(l)\) is verified, it is necessarily an innermost existential quantifier and a QCSP base containing only the values associated to the \BLTOP. is built thanks to the function \(base\_case\) defined by \(base\_case(l) = \{(val, \Box) | (val, \BLTOP.)\in l\}\).
Otherwise, the returned QCSP base is built by adding to the result of the \COMBINE. operator the list of the values associated to each of the QCSP bases thanks to the function \(first\_values\) defined by \(first\_values(l) = \{(val, \Box) | (val, a)\in l\}\).

\begin{myalgorithm}[h!]
\label{algo:combilexiste}
\caption{\COMBILEXISTE.}
\begin{mycode}
\ENTREE{\(x_{D}\) : a variable and its domain}
\ENTREE{\BINDER. : a binder}
\ENTREE{\(l\) : a list of pairs (value, QCSP base)}
\ENSURE{a QCSP base}
\IF{\(constants(l)\)}
 \RETURN{\(\bl{\ILEXISTE.x_{D}\BINDER.}{case\_base(l)}\)}
\ELSE
 \RETURN{\(\bl{\ILEXISTE.x_{D}\BINDER.}{[first\_values(l) | \combine{x}{l}]}\)}
\ENDIF
\end{mycode}
\end{myalgorithm}


\begin{myalgorithm}[h!]
\label{algo:combquelquesoit}
\caption{\COMBQUELQUESOIT.}
\begin{mycode}
\ENTREE{\(x_{D}\) : a variable and its domain}
\ENTREE{\BINDER. : a binder}
\ENTREE{\(l\) : a list of pairs (value, QCSP base)}
\ENSURE{a QCSP base or \BLTOP.}
\IF{\(constants(l)\)}
 \RETURN{\(\BLTOP.\)}
\ELSE
 \RETURN{\(\bl{\QUELQUESOIT.x_{D}\BINDER.}{\combine{x}{l}}\)}
\ENDIF
\end{mycode}
\end{myalgorithm}

The \COMBINE. operator works as follows.
The \(decompose\) function extracts from the list \(lvb\) of pairs (value, QCSP bases), for the outermost existentially quantified variable $y$ of the binder, a pair constituted of a list of pairs (val, list of guards)) and a list of pairs (val, remaining of the guards).
The \(compose\) function builds for $y$ its set of guards by distributing the trees for the different values.
Functions \(first\) and \(second\) give access to respectively the first and the second position of a pair.

\begin{myalgorithm}[h!]
\label{algo:combine}
\caption{\COMBINE.}
\begin{mycode}
\ENTREE{\(x\) : a variable}
\ENTREE{\(lvb\) : a list of pairs (value, QCSP base)}
\ENSURE{a list of guards}
\STATE{\(lg \leftarrow []\)}
\STATE{\(lvg \leftarrow extract\_guards(lvb)\)}
\WHILE{\(\vide{lvg}\)}
\STATE{\(dec\_y \leftarrow decompose(lvg)\)}
\STATE{\(lg \leftarrow [compose(x, first(dec\_y) | lg]\)}
\STATE{\(lvg \leftarrow second(dec\_y)\)}
\ENDWHILE
\RETURN{\(lg\)}
\end{mycode}
\end{myalgorithm}

The following example shows how the \RCQCSP. algorithm works.

\begin{example}
\label{exe:rec_comp}
We compute for all \(val_x,val_y\in \{0,1,2\}\) the QCSP base \(B_{val_x val_y}\) as the result of the following call~:
\[\RCQCSP.(\QUELQUESOIT. z \ILEXISTE. t,\{(x= (y*z)+t), (x=val_x),(y=val_y)\})\]
with \(z,t\in {\{0,1,2\}}\).

We obtain the QCSP bases (according to $T^{val_xval_y}_{val_t}$)~:

\[\begin{array}{l}
B_{00}=\langle
\QUELQUESOIT. z \ILEXISTE. t
|
\;\;
[
[(0, T^{00}_{0}= 
\tikzstyle{level 1}=[level distance=1cm, sibling distance=0.7cm]
\begin{tikzpicture}[grow = down]
\node [circle,draw] (root) {$z$}
child {node [draw] (x0y2) {}
  edge from parent node[left] {0}
}
child {node [draw] (x0y1) {}
  edge from parent node[left] {1}
}
child {node [draw] (x0y0) {}
  edge from parent node[right] {2}
};
\end{tikzpicture}
)
]
]
\rangle
\end{array}\]

\[\begin{array}{l}
B_{10}=\langle
\QUELQUESOIT. z \ILEXISTE. t
|
\;\;
[
[(1, T^{10}_{1}= 
\tikzstyle{level 1}=[level distance=1cm, sibling distance=0.7cm]
\begin{tikzpicture}[grow = down]
\node [circle,draw] (root) {$z$}
child {node [draw] (x0y2) {}
  edge from parent node[left] {0}
}
child {node [draw] (x0y1) {}
  edge from parent node[left] {1}
}
child {node [draw] (x0y0) {}
  edge from parent node[right] {2}
};
\end{tikzpicture}
)
]
]
\rangle
\end{array}\]

\[\begin{array}{l}
B_{20}=\langle
\QUELQUESOIT. z \ILEXISTE. t
|
\;\;
[
[(2, T^{20}_{2}= 
\tikzstyle{level 1}=[level distance=1cm, sibling distance=0.7cm]
\begin{tikzpicture}[grow = down]
\node [circle,draw] (root) {$z$}
child {node [draw] (x0y2) {}
  edge from parent node[left] {0}
}
child {node [draw] (x0y1) {}
  edge from parent node[left] {1}
}
child {node [draw] (x0y0) {}
  edge from parent node[right] {2}
};
\end{tikzpicture}
)
]
]
\rangle
\end{array}\]

\[B_{21}=\langle
\QUELQUESOIT. z \ILEXISTE. t \;
|
\;\;
{[}
[
(0, T^{21}_{0}=
\tikzstyle{level 1}=[level distance=1cm, sibling distance=0.7cm]
\begin{tikzpicture}[grow = down]
\node [circle,draw] (root) {$z$}
child {node [draw] (x0y2) {}
  edge from parent node[left] {2}
};
\end{tikzpicture}
),
(1, T^{21}_{1}= 
\tikzstyle{level 1}=[level distance=1cm, sibling distance=0.7cm]
\begin{tikzpicture}[grow = down]
\node [circle,draw] (root) {$z$}
child {node [draw] (x0y2) {}
  edge from parent node[left] {1}
};
\end{tikzpicture}
),
(2, T^{21}_{2}= 
\tikzstyle{level 1}=[level distance=1cm, sibling distance=0.7cm]
\begin{tikzpicture}[grow = down]
\node [circle,draw] (root) {$z$}
child {node [draw] (x0y2) {}
  edge from parent node[left] {0}
};
\end{tikzpicture}
)
]
]
\rangle
\]

of for any other combination, \(B_{val_x val_y}= \BLBOTTOM.\).

\end{example}

The following example shows how the trees are shared by the \COMBINE. operator and also the distribution of the trees.

\begin{example}
\label{exe:combilexiste}
The operator \COMBILEXISTE. is applied during the execution of the call
\[
\RCQCSP.(\ILEXISTE.y \QUELQUESOIT. z \ILEXISTE. t,
\{(x= (y*z)+t), (x=2)\})
\]
with \(y,z,t\in {\{0,1,2\}}\), to the QCSP bases \(B_{20}\), \(B_{21}\) and \(B_{22}\) which  represent compatible QCSP bases with QCSP, respectively, \({\QUELQUESOIT. z \ILEXISTE. t}{((x= (y*z)+t)\ET. (x=2)\ET. (y=0))}\), \({\QUELQUESOIT. z \ILEXISTE. t}{((x= (y*z)+t)\ET. (x=2) \ET. (y=1))}\), \({\QUELQUESOIT. z \ILEXISTE. t}{((x= (y*z)+t) \ET. (x=2) \ET. (y=2))}\).

\[\begin{array}{ll}
\lefteqn{\COMBILEXISTE.(y, \QUELQUESOIT. z \ILEXISTE. t,[(0,B_{20}),(1,B_{21}),(2,B_{22})])}
\\
\\
= & 
\langle
\ILEXISTE.y \QUELQUESOIT. z \ILEXISTE. t \;
| \;\; [[(0, \Box ),(1, \Box)],
[
(0, 
\tikzstyle{level 1}=[level distance=1cm, sibling distance=1cm]
\begin{tikzpicture}[grow = down]
\node [circle,draw] (root) {$y$}
child {node [] (x2t0y1) {$T^{21}_{0}$}
  edge from parent node[left] {1}
};
\end{tikzpicture}
),
(1, 
\tikzstyle{level 1}=[level distance=1cm, sibling distance=1cm]
\begin{tikzpicture}[grow = down]
\node [circle,draw] (root) {$y$}
child {node [] (x2t1y1) {$T^{21}_{1}$}
  edge from parent node[left] {1}
};
\end{tikzpicture}
),
(2, 
\tikzstyle{level 1}=[level distance=1cm, sibling distance=1cm]
\begin{tikzpicture}[grow = down]
\node [circle,draw] (root) {$y$}
child {node [] (x2t2y0) {$T^{20}_{2}$}
  edge from parent node[left] {0}
}
child {node [] (x2t2y1) {$T^{21}_{2}$}
  edge from parent node[right] {1}
};
\end{tikzpicture}
)
]
]
\rangle
\\
= & B_2
\end{array}\]

which is a compatible QCSP base with the QCSP \({\ILEXISTE.y \QUELQUESOIT. z \ILEXISTE. t}{((x= (y*z)+t) \ET. (x=2))}\)
with \(y,z,t\in {\{0,1,2\}}\).

\end{example}

The following example shows how the existential player can play with an optimal QCSP base instead of only one winning strategy how the optimal QCSP base gives a direct access to the possibilities for a given existentially quantified variable.

\begin{example}
\label{exe:follow_the_base}
{\em (Previous examples continued.)} Let the following QCSP be the expression of a very simple two-player game:
\[\ILEXISTE.x\ILEXISTE.y\QUELQUESOIT.z\ILEXISTE.t (x = (y*z)+t), x,y,z,t\in {\{0,1,2\}}\]

Let us suppose that an existential player only knows the following winning strategy:

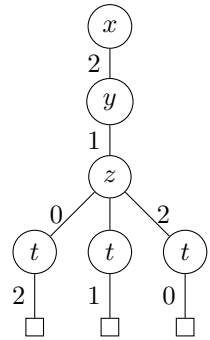
\begin{center}
{
\tikzstyle{level 1}=[level distance=1cm, sibling distance=1cm]
\tikzstyle{level 2}=[level distance=1cm, sibling distance=1cm]
\tikzstyle{level 3}=[level distance=1cm, sibling distance=1cm]
\tikzstyle{level 4}=[level distance=1cm, sibling distance=0.7cm]

\begin{tikzpicture}[grow = down]
\node [circle,draw] (root){$x$} 
child {node [circle,draw] (x0) {$y$}
  child {node [circle,draw] (x0y0) {$z$}
    child {node [circle,draw] (x0y0z2) {$t$}
      child {node [draw] (x0y0z2t0) {} edge from parent node[left] {2}}
      edge from parent node[left] {0}
    }
    child {node [circle,draw] (x0y0z1) {$t$}
      child {node [draw] (x0y0z1t0) {} edge from parent node[left] {1}}
    }   
    child {node [circle,draw] (x0y0z0) {$t$}
      child {node [draw] (x0y0z0t0) {} edge from parent node[left] {0}}
      edge from parent node[right] {2}
    }
    edge from parent node[left] {1}
  }
  edge from parent node[left] {2}
};
\end{tikzpicture}
}
\end{center}

For his first move, he decides to play \((x=2)\).
Now, following its winning strategy, he is supposed to play \((y=1)\).
Let us suppose that it is no more possible because of an unexpected reason.
He can not follow its winning strategy anymore.
If he follows the compatible but not optimal base of Example~\ref{exe:base_EEAE} he will follows one of the two other choices thinking that he still has a chance.
If he wants to be sure, he will have to pay the full computational price for the QCSP 
\(\QUELQUESOIT.z\ILEXISTE.t ((x = (y*z)+t) \ET. (x=2) \ET. (y=0)), z,t\in {\{0,1,2\}}\)
and 
\(\QUELQUESOIT.z\ILEXISTE.t ((x = (y*z)+t) \ET. (x=2) \ET. (y=2)), z,t\in {\{0,1,2\}}\).
If he follows its compatible and optimal QCSP base of Example~\ref{exe:base_EEAE_opt}, he knows that he has already lost.

\end{example}


The following theorem establishes that the \RCQCSP. algorithm not only computes  a compatible QCSP base from a QCSP but also that this QCSP base is optimal.

\begin{theorem}
\label{th:optimalite_recursive}
Let \({\BINDER.}{C}\) be a QCSP. 
\(\rcqcsp{\BINDER.}{C}\) return an optimal base and compatible with the QCSP.

\end{theorem}

\section{Conclusion}
\label{sec:conclusion}
We have proposed in this article a framework for the compilation of QCSP: the QCSP bases.
We have defined an algorithm embedded in the state-of-the-art search algorithm of QCSP solver to compute a QCSP base from a QCSP.
We have shown that the obtained QCSP base is compatible with the initial QCSP and optimal in the sense that the construction of any winning strategy is polytime for the interpreration of a QCSP base.

We have implemented in Prolog the algorithms described in this article and the programs and examples are downloadable.
 at the following address
\url{http://www.info.univ-angers.fr/pub/stephan/Research/Download.html}.
We plan to integrate it in our QCSP solveur developed in the generic constraint development environment Gecode~\cite{Schulte_Tack_CP_05}.

When a QCSP solver returns there is or there is no winning strategy, there is no way to check if the answer is correct while for CSP the associated result to the decision (a complete instantiation) is easy to check. 
A certificate for a QCSP which has a winning strategy is any piece of information that provides self-supporting evidence of the existence of a winning strategy for that QCSP.
Due to the lack of space, we have not treated certificates for QCSP: our formalism  includes QCSP certificates as a particular case.
During the execution of the solver, a QCSP base is generated but only with strategies as guards.
The interpretation of these trees (cf~Definition~\ref{def:inter_arbre}) in tuples permits to verify such a QCSP certificate w.r.t. a QCSP by the resolution of a co-NP-complete problem. (The complexity is similar to the check of a winning strategy for QBF~\cite{Benedetti_IJCAI_05}.)

We have proposed a recursive construction of the QCSP base but in practice it is often more efficient to consider a cooperation between a solver which emits a trace and a trace analyzer which builds the QCSP base. 
We have also develop this approach but due to the lack of space we only give here the main two important reasons for this cooperation between a solver and a trace analyzer:
If the construction of the QCSP base is embedded into the solver, the memory management of the solver by means of a backtrack stack will have also to keep the state of the current QCSP base.
We want to take into account modern architectures with  multi-core multithreaded processors.

\bibliography{references}
\end{document}